\documentclass[nonacm]{acmart}

\usepackage[utf8]{inputenc}
\usepackage{array,ragged2e}

\newcolumntype{L}[1]{>{\RaggedRight}p{#1}}  
\newcolumntype{R}[1]{>{\RaggedLeft}p{#1}}   
\newcolumntype{C}[1]{>{\Centering}p{#1}}    
\newcolumntype{M}[1]{>{\RaggedRight}m{#1}}  
\newcolumntype{B}[1]{>{\RaggedRight}b{#1}}  

\AtBeginDocument{
  \providecommand\BibTeX{{
    \normalfont B\kern-0.5em{\scshape i\kern-0.25em b}\kern-0.8em\TeX}}}

\acmBooktitle{arXiv}

\begin{document}

\title[Survey about protection motivation on social networking sites]{Survey about protection motivation on social networking sites: University of Maribor students, 2018}

\author{Luka Jelovčan}
\email{luka.jelovcan@student.um.si}
\affiliation{
  \institution{University of Maribor}
  \country{Slovenia}
}

\author{Simon Vrhovec}
\email{simon.vrhovec@um.si}
\affiliation{
  \institution{University of Maribor}
  \country{Slovenia}
}

\author{Damjan Fujs}
\email{damjan.fujs@fri.uni-lj.si}
\affiliation{
  \institution{University of Ljubljana and University of Maribor}
  \country{Slovenia}
}

\begin{abstract}
This paper reports on a study aiming to explore factors associated with protection motivation of users on social networking sites. The objectives of this study were to determine how trust in internet service provider, trust in social media provider, trust in government, privacy concerns, fear of government intrusions, locus of control, and perceived threats affect protection motivation of users on social networking sites. The study employed a cross-sectional research design. A survey was conducted among University of Maribor students between October 2018 and January 2019. A total of 289 respondents completed the survey providing for N=276 useful responses after excluding poorly completed responses (27.9 percent response rate). The survey questionnaire was developed in English. A Slovenian translation of the survey questionnaire is available.
\end{abstract}

\keywords{PMT, social network, social media, cybersecurity, cyber threat, computer security, internet}

\maketitle

\section{Introduction}

The survey questionnaire was designed to measure theoretical constructs included in the research model. All items were taken or adapted from existing literature to fit the study's context. Table \ref{table:constructs} presents the theoretical constructs included in the research model, their definition in this research, and sources from which construct items were taken or adapted.

\section{Method}

\subsection{Survey instrument}

To test the research model a survey questionnaire was developed. Adapted questionnaire items (i.e., trust in internet service provider, trust in social media provider, trust in government, fear of government intrusions, locus of control, and political orientation) were developed by following a predefined protocol. The questionnaire was first developed in English and then translated into Slovenian by two translators independently. The translators developed the Slovenian questionnaire through consensus. The Slovenian questionnaire has been pre-tested by 3 independent respondents who provided feedback on its clarity. Based on the received feedback, the Slovenian questionnaire was reviewed to remove any ambiguity. Items were reworded, added, and deleted in the pre-test. To ensure the consistency between the Slovenian and English questionnaire, the Slovenian questionnaire was translated back to English. No significant differences in the meaning between the original items in English and back-translations were noticed. The English questionnaire was however reviewed to update the items and to remove any ambiguity based on the back-translation.

\clearpage

\begin{table}[h!]
\caption{\label{table:constructs}Theoretical constructs in the survey questionnaire.}
\small
\begin{tabular}{L{.24\textwidth}L{.51\textwidth}L{.18\textwidth}}
\toprule
Theoretical construct & Definition in this research & Sources \\
\midrule
Trust in internet service provider & The degree of trust in internet service provider. & Trusting beliefs \cite{McKnight2002} \\
Trust in social media provider & The degree of trust in social media provider. & Trusting beliefs \cite{McKnight2002} \\
Trust in government & The degree of trust in government of the current country of residence. & Trusting beliefs \cite{McKnight2002}\\
Privacy concerns & The extent of concerns regarding privacy on social networking sites. & Privacy concerns \cite{Fujs2019} \\
Fear of government intrusions & The extent of fear regarding government intrusions into privacy. & Fear \cite{Osman1994} \\
Locus of control & The extent to which individuals take responsibility for self-protecting on social networking sites themselves. & Locus of control \cite{Jansen2018} \\
Perceived threats & The degree to which intrusions into accounts on social networking sites threaten their users. & Perceived threats \cite{Fujs2019} \\
Protection motivation & The degree of motivation to self-protect on social networking sites. & Behavioral intent \cite{Fujs2019} \\
Political orientation & The extent to which individual tag themselves as "left" or "right". & Political orientation \cite{Kroh2007} \\
\bottomrule
\end{tabular}
\end{table}

Table \ref{table:questionnaire_en} presents the survey questionnaire in English and Table \ref{table:questionnaire_si} presents the Slovenian translation of the survey questionnaire. Most items were measured using a 5-point Likert scale as presented in Table \ref{table:likert}. Political orientation was measured with a 11-point scale as presented in Table \ref{table:po-scale}.

\subsection{Data collection}

We conducted the survey with the Slovenian translation of the questionnaire among University of Maribor students between 29 October 2018 and 28 January 2019. A total of 289 respondents completed the survey. After excluding poorly completed responses (responses with over 50 percent of missing values or standard deviation equal to 0 for constructs trust in internet service provider, trust in social media provider, trust in government, privacy concerns, fear of government intrusions, locus of control, perceived threats, and protection motivation), we were left with 276 useful responses providing for a response rate of 27.9 percent as presented in Table \ref{table:sample}.

\begin{table}[h!]
\caption{\label{table:sample}Sample with the size of the population, number of responses, and number of useful responses ($N$) after excluding poorly completed responses.}
\small
\begin{tabular}{L{.02\textwidth}L{.22\textwidth}R{.05\textwidth}R{.12\textwidth}R{.05\textwidth}}
\toprule
ID & Name & Size & Responses & $N$ \\
\midrule
1 & University of Maribor & 988 & 289 & 276 \\
\bottomrule
\end{tabular}
\end{table}

Due to the sensitive nature of the survey topic, safeguards were put in place to encourage participation and honest responses. First, the respondents were informed about the voluntariness and anonymity of participating in the survey. Next, the respondents were assured that the collected data will be used for research purposes only. No special incentives were offered to encourage participation in the survey.

\clearpage

\begin{table}[h!]
\caption{\label{table:questionnaire_en}Survey questionnaire items (English original).}
\small
\begin{tabular}{L{.35\textwidth}L{.60\textwidth}}
\toprule
Construct & Prompt/Item \\
\midrule
Trust in internet service provider (TiISP) & Mark your agreement with statements about your internet service provider: \\
 & TiISP1. I believe that the internet service provider would act in my best interest. \\
 & TiISP2. The internet service provider is interested in my well-being not just its own. \\
 & TiISP3. I would characterize the internet service provider as honest. \\

Trust in social media provider (TiSMP) & Mark your agreement with statements about providers of social networks on which you have accounts (e.g., Facebook, Twitter, Instragram, Snapchat, WhatsApp, Telegram, Tinder): \\
 & TiSMP1. I believe that social network providers would act in my best interest. \\
 & TiSMP2. Social network providers are interested in my well-being not just their own. \\
 & TiSMP3. I would characterize social network providers as honest. \\

Trust in government (TiG) & Mark your agreement with statements about the government of the country where you currently reside: \\
 & TiG1. I believe that the government would act in my best interest. \\
 & TiG2. The government is interested in my well-being not just its own. \\
 & TiG3. I would characterize the government as honest. \\

Privacy concerns (PC) & Mark your agreement with statements about your personal data on social networks (e.g., Facebook, Twitter, Instragram, Snapchat, WhatsApp, Telegram, Tinder): \\
 & PC1. It highly bothers me when social networks ask me about my personal data. \\
 & PC2. I always think twice before submitting my personal data to social networks. \\
 & PC3. I am very concerned that social networks collect too much personal data about me. \\

Fear of government intrusions (FoGI) & Mark your agreement with statements about the government of the country where you currently reside: \\
 & FoGI1. Government intrusions into my privacy are terrifying. \\
 & FoGI2. I am afraid of government intrusions into my privacy. \\
 & FoGI3. The government might be seriously invading my privacy. \\

Locus of control (LoC) & Mark your agreement with statements about control over your social network accounts: \\
 & LoC1. Keeping my accounts safe is within my control. \\
 & LoC2. I believe that it is within my control to protect myself against hacking into my accounts. \\
 & LoC3. The primary responsibility for protecting my accounts against hacking belongs to me. \\

Perceived threats (PT) & Mark your agreement with statements about potential intrusions into one of your social network accounts: \\
 & PT1. I feel threatened by intrusions. \\
 & PT2. Intrusions threaten my accounts. \\
 & PT3. It would be dreadful if there would be an intrusion into one of my accounts. \\

Protection motivation (PM) & Mark your agreement with statements about implementing recommended security measures on social networks (e.g., periodical password changes, use of strong passwords, paying attention to login alerts): \\
 & PM1. I intend to implement recommended security measures regularly. \\
 & PM2. I predict that I will implement recommended security measures in the near future. \\
 & PM3. I plan to implement recommended security measures. \\

Political orientation (PO) & In politics, people sometimes talk of ‘left’ and ‘right’. Where would you place yourself on the scale below? \\
\bottomrule
\end{tabular}
\end{table}

\begin{table}[h!]
\caption{\label{table:questionnaire_si}Survey questionnaire items (Slovenian translation).}
\small
\begin{tabular}{L{.35\textwidth}L{.60\textwidth}}
\toprule
Construct & Prompt/Item \\
\midrule
Trust in internet service provider (TiISP) & Označite svoje strinjanje s trditvami o ponudniku vaših internetnih storitev: \\
 & TiISP1. Verjamem, da ponudnik internetnih storitev deluje v mojem najboljšem interesu. \\
 & TiISP2. Ponudnika internetnih storitev zanima tudi moja blaginja, ne le njegova. \\
 & TiISP3. Ponudnika internetnih storitev bi opredelil kot poštenega. \\

Trust in social media provider (TiSMP) & Označite svoje strinjanje s trditvami o ponudnikih socialnih omrežij, na katerih imate račune (npr. Facebook, Twitter, Instagram, Snapchat, Whats App, Telegram, Tinder): \\
 & TiSMP1. Verjamem, da ponudniki socialnih omrežij delujejo v mojem najboljšem interesu. \\
 & TiSMP2. Ponudnike socialnih omrežij zanima tudi moja blaginja, ne le njihova. \\
 & TiSMP3. Ponudnike socialnih omrežij bi opredelil kot poštene. \\

Trust in government (TiG) & Označite svoje strinjanje s trditvami o vladi države, v kateri trenutno bivate: \\
 & TiG1. Verjamem, da vlada deluje v mojem najboljšem interesu. \\
 & TiG2. Vlado zanima tudi moja blaginja, ne le njena. \\
 & TiG3. Vlado bi opredelil kot pošteno. \\

Privacy concerns (PC) & Označite svoje strinjanje s trditvami o vaših osebnih podatkih na socialnih omrežjih (npr. Facebook, Twitter, Instagram, Snapchat, Whats App, Telegram, Tinder): \\
 & PC1. Zelo me moti, ko me socialna omrežja sprašujejo po osebnih podatkih. \\
 & PC2. Preden posredujem svoje osebne podatke socialnim omrežjem, vedno premislim dvakrat. \\
 & PC3. Zelo me skrbi, da socialna omrežja o meni zbirajo preveč osebnih podatkov. \\

Fear of government intrusions (FoGI) & Označite svoje strinjanje s trditvami o vladi države, v kateri trenutno bivate: \\
 & FoGI1. Možni vdori vlade v mojo zasebnost so zastrašujoči. \\
 & FoGI2. Bojim se vdorov vlade v mojo zasebnost. \\
 & FoGI3. Vlada lahko resno ogrozi mojo zasebnost. \\

Locus of control (LoC) & Označite svoje strinjanje s trditvami o nadzoru nad vašimi računi na socialnih omrežjih: \\
 & LoC1. Varovanje mojih računov je pod mojim nadzorom. \\
 & LoC2. Verjamem, da je zaščita pred vdori v moje račune pod mojim nadzorom. \\
 & LoC3. Primarna odgovornost za zaščito pred vdori v moje račune je moja. \\

Perceived threats (PT) & Označite svoje strinjanje s trditvami o morebitnih vdorih v katerega izmed vaših računov na socialnih omrežjih: \\
 & PT1. Zaradi vdorov se počutim ogroženega. \\
 & PT2. Vdori ogrožajo moje račune. \\
 & PT3. Grozljivo bi bilo, če bi vdrli v katerega izmed mojih računov. \\

Protection motivation (PM) & Označite svoje strinjanje s trditvami o uporabi priporočenih varnostnih mehanizmov na socialnih omrežjih (npr. redno spreminjanje gesel, uporaba močnih gesel, posvečanje pozornosti opozorilom o prijavi na socialno omrežje): \\
 & PM1. Redno nameravam uporabljati priporočene varnostne mehanizme. \\
 & PM2. Predvidevam, da bom v bližnji prihodnosti uporabljal priporočene varnostne mehanizme. \\
 & PM3. Načrtujem uporabo priporočenih varnostnih mehanizmov. \\

Political orientation (PO) & V politiki ljudje včasih govorijo o "levici" in "desnici". Kako bi se opredelili na spodnji lestvici? \\
\bottomrule
\end{tabular}
\end{table}

\clearpage

\begin{table}[h!]
\caption{\label{table:likert}5-point Likert scale.}
\small
\begin{tabular}{L{.06\textwidth}L{.19\textwidth}L{.19\textwidth}}
\toprule
Score & English & Slovenian \\
\midrule
1 & Strongly disagree & Močno se ne strinjam \\
2 & Disagree & Se ne strinjam \\
3 & Neutral & Nevtralno \\
4 & Agree & Se strinjam \\
5 & Strongly agree & Močno se strinjam \\
\bottomrule
\end{tabular}
\end{table}

\begin{table}[h!]
\caption{\label{table:po-scale}11-point political orientation scale.}
\small
\begin{tabular}{L{.06\textwidth}L{.19\textwidth}L{.19\textwidth}}
\toprule
Score & English & Slovenian \\
\midrule
1 & Completely left & Skrajno levo \\
2 & & \\
3 & & \\
4 & & \\
5 & & \\
6 & Neutral & Neopredeljen \\
7 & & \\
8 & & \\
9 & & \\
10 & & \\
11 & Completely right & Skrajno desno \\
\bottomrule
\end{tabular}
\end{table}

The first page of the survey is presented in Table \ref{table:first-page}.

\begin{table}[h!]
\caption{\label{table:first-page}The first page of the survey.}
\small
\begin{tabular}{L{.47\textwidth}L{.47\textwidth}}
\toprule
English original & Slovenian translation \\
\midrule
\textit{Social networks} \newline ~ \newline Dear Sirs, \newline ~ \newline In front of you is a survey on the security of users on social networks (e.g., Facebook, Twitter, Instagram, Snapchat, WhatsApp, Telegram, Tinder) conducted at the Faculty of Criminal Justice and Security at the University of Maribor. Estimated time to complete the survey is 2--3 minutes. \newline ~ \newline Participation in the research is voluntary and anonymous while the data will be used exclusively for research purposes. \newline ~ \newline If you are taking the survey via a smartphone, note that each set has 3 questions. After selecting an answer, the next question in the set is automatically displayed. Tip: Turn the smartphone horizontally. \newline ~ \newline Contact: \newline Damjan Fujs (damjan.fujs@student.um.si) & \textit{Socialna omrežja} \newline ~ \newline Spoštovani, \newline ~ \newline pred vami je raziskava o varnosti uporabnikov na socialnih omrežjih (npr. Facebook, Twitter, Instagram, Snapchat, Whats App, Telegram, Tinder), ki jo izvajamo na Fakulteti za varnostne vede Univerze v Mariboru. Predviden čas izpolnjevanja ankete je 2-3 minute. \newline ~ \newline Sodelovanje v raziskavi je prostovoljno in anonimno, podatki pa bodo uporabljeni izključno za raziskovalne namene. \newline ~ \newline V kolikor anketo izpolnjujete preko pametnega telefona, bodite pozorni na to, da ima vsak sklop 3 vprašanja. Ob izbiri odgovora se samodejno prikaže naslednje vprašanje v sklopu. Namig: Pametni telefon obrnite vodoravno. \newline ~ \newline Kontakt: \newline Damjan Fujs (damjan.fujs@student.um.si) \\
\bottomrule
\end{tabular}
\end{table}

\section{Results}

\subsection{Sample}

Demographic characteristics of the sample are presented in Table \ref{table:demographics}.

\begin{table}[h!]
\caption{\label{table:demographics}Demographic characteristics of the sample.}
\small
\begin{tabular}{L{.18\textwidth}L{.24\textwidth}R{.08\textwidth}}
\toprule
Characteristic & Value & Frequency \\
\midrule
Gender & 1 -- Male & 91 \\
 & 2 -- Female & 176 \\
 & \textit{Missing} & 22 \\
 
Age & 18 & 1 \\
 & 19 & 35 \\
 & 20 & 32 \\
 & 21 & 33 \\
 & 22 & 36 \\
 & 23 & 34 \\
 & 24 & 23 \\
 & 25 & 26 \\
 & 26 & 4 \\
 & 27 & 4 \\
 & 28 & 2 \\
 & 29 & 5 \\
 & 30 & 5 \\
 & 31 & 2 \\
 & 32 & 3 \\
 & 33 & 1 \\
 & 34 & 1 \\
 & 35 & 4 \\
 & 36 & 2 \\
 & 37 & 3 \\
 & 38 & 2 \\
 & 40 & 3 \\
 & 42 & 2 \\
 & 43 & 1 \\
 & 44 & 1 \\
 & 49 & 1 \\
 & 53 & 1 \\
 & \textit{Missing} & 22 \\
 
Formal education & 1 -- Less than Bachelor's degree & 145 \\
 & 2 -- Bachelor's degree & 91 \\
 & 3 -- Master's degree & 25 \\
 & 4 -- PhD & 4 \\
 & \textit{Missing} & 24 \\
 
Employment status & 1 -- Student & 216 \\
 & 2 -- Employed & 41 \\
 & 3 -- Unemployed & 8 \\
 & \textit{Missing} & 24 \\
\bottomrule
\end{tabular}
\end{table}

\clearpage

\subsection{Frequencies}

Frequencies of most used social media are presented in Table \ref{table:social_media}.

\begin{table}[h!]
\caption{\label{table:social_media}Frequencies of most used social media.}
\small
\begin{tabular}{L{.24\textwidth}R{.08\textwidth}}
\toprule
Most used social media & Frequency \\
\midrule
1 -- Facebook & 146 \\
2 -- WhatsApp & 2 \\
7 -- Instagram & 80 \\
8 -- Twitter & 6 \\
9 -- Google+ & 3 \\
11 -- Viber & 7 \\
12 -- Shina Weibo & 1 \\
13 -- Snapchat & 9 \\
15 -- Pinterest & 1 \\
16 -- LinkedIn & 1 \\
17 -- Telegram & 1 \\
18 -- Reddit & 4 \\
23 -- Other: Youtube & 1 \\
\textit{Missing} & 27 \\
\bottomrule
\end{tabular}
\end{table}

Political orientation of the sample is presented in Table \ref{table:political_orientation}.

\begin{table}[h!]
\caption{\label{table:political_orientation}Political orientation of the sample.}
\small
\begin{tabular}{L{.24\textwidth}R{.08\textwidth}}
\toprule
Political orientation & Frequency \\
\midrule
1 & 8 \\
2 & 5 \\
3 & 12 \\
4 & 22 \\
5 & 21 \\
6 & 165 \\
7 & 8 \\
8 & 15 \\
9 & 8 \\
10 & 0 \\
11 & 3 \\
\textit{Missing} & 22 \\
\bottomrule
\end{tabular}
\end{table}

\clearpage

Frequencies of all variables for measured theoretical constructs are presented in Table \ref{table:Frequencies}.

\begin{table}[h!]
\caption{\label{table:Frequencies}Frequencies of variables.}
\small
\begin{tabular}{L{.07\textwidth}R{.03\textwidth}R{.03\textwidth}R{.03\textwidth}R{.03\textwidth}R{.03\textwidth}R{.07\textwidth}R{.07\textwidth}R{.07\textwidth}}
\toprule
Variable & 1 & 2 & 3 & 4 & 5 & Valid & Missing & Total \\
\midrule
TiISP1 & 11 & 81 & 114 & 73 & 9 & 288 & 1 & 289 \\
TiISP2 & 19 & 92 & 100 & 64 & 13 & 288 & 1 & 289 \\
TiISP3 & 16 & 70 & 136 & 59 & 4 & 284 & 5 & 289 \\

TiSMP1 & 33 & 102 & 93 & 54 & 6 & 288 & 1 & 289 \\
TiSMP2 & 31 & 95 & 88 & 64 & 8 & 286 & 3 & 289 \\
TiSMP3 & 34 & 104 & 110 & 36 & 1& 285 & 4 & 289 \\

TiG1 & 41 & 109 & 97 & 38 & 2 & 287 & 2 & 289 \\
TiG2 & 45 & 91 & 80 & 56 & 11 & 283 & 6 & 289 \\
TiG3 & 79 & 113 & 78 & 13 & 2 & 285 & 4 & 289 \\

PC1 & 2 & 26 & 37 & 131 & 81 & 277 & 12 & 289 \\
PC2 & 4 & 16 & 21 & 134 & 102 & 277 & 12 & 289 \\
PC3 & 1 & 13 & 26 & 109 & 127 & 276 & 13 & 289 \\

FoGI1 & 5 & 42 & 68 & 115 & 46 & 276 & 13 & 289 \\
FoGI2 & 12 & 59 & 73 & 93 & 39 & 276 & 13 & 289 \\
FoGI3 & 8 & 37 & 91 & 89 & 47 & 272 & 17 & 289 \\

LoC1 & 7 & 57 & 91 & 92 & 21 & 268 & 21 & 289 \\
LoC2 & 12 & 81 & 98 & 66 & 10 & 267 & 22 & 289 \\
LoC3 & 3 & 33 & 71 & 126 & 34 & 267 & 22 & 289 \\

PT1 & 5 & 44 & 77 & 115 & 27 & 268 & 21 & 289 \\
PT2 & 4 & 36 & 88 & 116 & 23 & 267 & 22 & 289 \\
PT3 & 2 & 17 & 34 & 115 & 98 & 266 & 23 & 289 \\

PM1 & 3 & 15 & 57 & 153 & 40 & 268 & 21 & 289 \\
PM2 & 2 & 11 & 44 & 165 & 45 & 267 & 22 & 289 \\
PM3 & 2 & 13 & 51 & 155 & 43 & 264 & 25 & 289 \\
\bottomrule
\end{tabular}
\end{table}

\section{Discussion}

This paper presents the results of a survey about protection motivation on social networking sites. Future studies may focus on other factors associated with protection motivation on social networking sites. Such studies would be beneficial both to better explain protection motivation of users on social networking sites, and to better understand the associations between different factors associated with it.

\bibliographystyle{ACM-Reference-Format}
\bibliography{main}


\begin{thebibliography}{5}


\ifx \showCODEN    \undefined \def \showCODEN     #1{\unskip}     \fi
\ifx \showDOI      \undefined \def \showDOI       #1{#1}\fi
\ifx \showISBNx    \undefined \def \showISBNx     #1{\unskip}     \fi
\ifx \showISBNxiii \undefined \def \showISBNxiii  #1{\unskip}     \fi
\ifx \showISSN     \undefined \def \showISSN      #1{\unskip}     \fi
\ifx \showLCCN     \undefined \def \showLCCN      #1{\unskip}     \fi
\ifx \shownote     \undefined \def \shownote      #1{#1}          \fi
\ifx \showarticletitle \undefined \def \showarticletitle #1{#1}   \fi
\ifx \showURL      \undefined \def \showURL       {\relax}        \fi
\providecommand\bibfield[2]{#2}
\providecommand\bibinfo[2]{#2}
\providecommand\natexlab[1]{#1}
\providecommand\showeprint[2][]{arXiv:#2}

\bibitem[\protect\citeauthoryear{Fujs, Miheli{\v{c}}, and Vrhovec}{Fujs
  et~al\mbox{.}}{2019}]%
        {Fujs2019}
\bibfield{author}{\bibinfo{person}{Damjan Fujs}, \bibinfo{person}{An{\v{z}}e
  Miheli{\v{c}}}, {and} \bibinfo{person}{Simon Vrhovec}.}
  \bibinfo{year}{2019}\natexlab{}.
\newblock \showarticletitle{{Social Network Self-Protection Model: What
  Motivates Users to Self-Protect?}}
\newblock \bibinfo{journal}{\emph{Journal of Cyber Security and Mobility}}
  \bibinfo{volume}{8}, \bibinfo{number}{4} (\bibinfo{year}{2019}),
  \bibinfo{pages}{467--492}.
\newblock
\urldef\tempurl%
\url{https://doi.org/10.13052/jcsm2245-1439.844}
\showDOI{\tempurl}


\bibitem[\protect\citeauthoryear{Jansen and van Schaik}{Jansen and van
  Schaik}{2018}]%
        {Jansen2018}
\bibfield{author}{\bibinfo{person}{Jurjen Jansen} {and} \bibinfo{person}{Paul
  van Schaik}.} \bibinfo{year}{2018}\natexlab{}.
\newblock \showarticletitle{{Testing a model of precautionary online behaviour:
  The case of online banking}}.
\newblock \bibinfo{journal}{\emph{Computers in Human Behavior}}
  \bibinfo{volume}{87} (\bibinfo{year}{2018}), \bibinfo{pages}{371--383}.
\newblock
\urldef\tempurl%
\url{https://doi.org/10.1016/j.chb.2018.05.010}
\showDOI{\tempurl}


\bibitem[\protect\citeauthoryear{Kroh}{Kroh}{2007}]%
        {Kroh2007}
\bibfield{author}{\bibinfo{person}{Martin Kroh}.}
  \bibinfo{year}{2007}\natexlab{}.
\newblock \showarticletitle{{Measuring Left-Right Political Orientation: The
  Choice of Response Format}}.
\newblock \bibinfo{journal}{\emph{Public Opinion Quarterly}}
  \bibinfo{volume}{71}, \bibinfo{number}{2} (\bibinfo{year}{2007}),
  \bibinfo{pages}{204--220}.
\newblock
\urldef\tempurl%
\url{https://doi.org/10.1093/poq/nfm009}
\showDOI{\tempurl}


\bibitem[\protect\citeauthoryear{McKnight, Choudhury, and Kacmar}{McKnight
  et~al\mbox{.}}{2002}]%
        {McKnight2002}
\bibfield{author}{\bibinfo{person}{D.~Harrison McKnight},
  \bibinfo{person}{Vivek Choudhury}, {and} \bibinfo{person}{Charles Kacmar}.}
  \bibinfo{year}{2002}\natexlab{}.
\newblock \showarticletitle{{The impact of initial consumer trust on intentions
  to transact with a web site: a trust building model}}.
\newblock \bibinfo{journal}{\emph{The Journal of Strategic Information
  Systems}} \bibinfo{volume}{11}, \bibinfo{number}{3-4} (\bibinfo{year}{2002}),
  \bibinfo{pages}{297--323}.
\newblock
\urldef\tempurl%
\url{https://doi.org/10.1016/S0963-8687(02)00020-3}
\showDOI{\tempurl}


\bibitem[\protect\citeauthoryear{Osman, Barrios, Osman, Schneekloth, and
  Troutman}{Osman et~al\mbox{.}}{1994}]%
        {Osman1994}
\bibfield{author}{\bibinfo{person}{Augustine Osman},
  \bibinfo{person}{Francisco~X. Barrios}, \bibinfo{person}{Joylene~R. Osman},
  \bibinfo{person}{Raelynn Schneekloth}, {and} \bibinfo{person}{Josh~A.
  Troutman}.} \bibinfo{year}{1994}\natexlab{}.
\newblock \showarticletitle{{The Pain Anxiety Symptoms Scale: Psychometric
  properties in a community sample}}.
\newblock \bibinfo{journal}{\emph{Journal of Behavioral Medicine}}
  \bibinfo{volume}{17}, \bibinfo{number}{5} (\bibinfo{year}{1994}),
  \bibinfo{pages}{511--522}.
\newblock
\urldef\tempurl%
\url{https://doi.org/10.1007/BF01857923}
\showDOI{\tempurl}


\end{thebibliography}

\end{document}